\documentstyle[prd,preprint,aps]{revtex} 
\begin{document}

\reversemarginpar
\tighten

\title{Symmetries at stationary Killing horizons}
\author{A.J.M.  Medved} 
\address{
School of Mathematics, Statistics and Computer Science\\
Victoria University of Wellington\\ 
PO Box 600, Wellington New Zealand \\
E-Mail: joey.medved@mcs.vuw.ac.nz}
\maketitle

\begin{abstract}

   It has often been suggested (especially by Carlip) that 
   spacetime symmetries in the neighborhood of a black hole horizon  
   may be relevant to a statistical understanding of the 
   Bekenstein--Hawking  entropy.  A prime candidate for this type of  
   symmetry is that  which is exhibited by the Einstein tensor.
   More precisely, it is now known that this tensor  takes on 
   a  strongly constrained (block-diagonal) form as it approaches any 
   stationary, non-extremal Killing horizon.  Presently, 
   exploiting the geometrical properties of such horizons, we provide 
   a particularly elegant argument that substantiates 
   this highly symmetric form for the Einstein tensor.    
    It is, however, duly noted that, on account of   a ``loophole'', 
    the argument does fall just short of attaining the status of 
    a rigorous proof.




\end{abstract}


\section{The Motivation}

No one seriously disputes the notion of black holes as thermodynamic entities; 
nevertheless,
the Bekenstein--Hawking entropy  \cite{BEK,HAW} remains
as enigmatic as ever from a statistical viewpoint.
The ``company line'' has been, more often than not,   to 
hope  that quantum gravity will eventually provide the resolution; but   
this could require, cynically speaking,
a rather long wait. A more pragmatic expectation
might be to hope for a statistical  explanation that  interpolates between
the quantum-gravitational and semi-classical regimes, 
and that is not particularly
sensitive to the fundamental micro-constituents.
Such a perspective appears to be in compliance with the general stance
of S.  Carlip --- who  has  long advocated
for horizon boundary conditions as a means of
altering the physical content of the theory, thereby  
inducing new degrees 
freedom that can account for the black hole entropy \cite{CAR6}.

One might then query as to what physical principle  
determines the correct choice of boundary conditions.
On this point, Carlip has stressed the importance
of asymptotic symmetries~\footnote{Asymptotic in the sense that
such symmetries need only be exact at the horizon itself.}   
in the neighborhood of the horizon \cite{CAR3}.
Ideally, these symmetries should be based on  semi-classical concepts that 
can be enhanced
into a quantum environment and that are 
strong enough to physically constrain the theory
so that it  describes only  black holes. Indeed, the merit
of this philosophy has been substantiated by the various statistical
calculations of 
Carlip \cite{CAR1,CARX} and others ({\it e.g.}, 
\cite{SOL,Z1,Z2,Z3,Z4,Z5,Z6,Z7}).
Remarkably, these studies have successfully replicated the Bekenstein--Hawking
entropy (although sometimes up to a constant factor) while following, 
more or less,  along the stated   
lines.~\footnote{Interestingly, the seminal calculation
of this nature \cite{STR} was based rather on 
asymptotic symmetries at spatial infinity \cite{BH}.}

One can find a viable candidate for Carlip's notion of an asymptotic symmetry
 by looking directly at the  near-horizon geometry
of a stationary (but otherwise arbitrary) black hole spacetime. To elaborate,
the current author, M. Visser and D. Martin have demonstrated
a highly constrained and symmetric form for the Einstein tensor 
near the horizon of any stationary (non-extremal) Killing horizon.
Irregardless of whether the horizon is static \cite{MMV} or
rotating (but still stationary) \cite{Prior}, we have shown that
the on-horizon Einstein tensor block-diagonalizes into
transverse ($\perp$) and parallel ($\parallel$) blocks.~\footnote{Orientations
are always  defined relative to an arbitrary spacelike cross-section
of the horizon. Also note that a four-dimensional (black hole) spacetime 
is assumed throughout.}
It was also demonstrated that the transverse block
must be
directly proportional to the (induced) transverse 
metric.~\footnote{This form had actually been known 
for quite some time, but only under
the stringent conditions of a  static {\it and} spherically symmetric 
horizon geometry \cite{DBH1}.}

To better appreciate these findings (especially in the context
of Carlip's ideas), let us make a couple of pertinent comments.
Firstly,  the block-diagonalization indicates that, at the horizon, 
the ``$r$--$t$''-plane dissociates from the in-horizon
coordinates, which is a necessary ingredient in most
statistical entropy calculations \cite{CAR3}.
Secondly, given the Einstein field equations,
it immediately follows that the on-horizon stress tensor is subject to
an identical set of constraints. As a consequence of the form
of the transverse block,  one
finds that (as the horizon is approached) the transverse pressure goes 
as  the negative of the energy density.
This, in turn, suggests
that the matter near a horizon can be effectively described
as a two-dimensional conformal field theory (see \cite{MMV,Prior}
for an explanation); a theory that necessarily underlies
any entropy  calculation that calls upon the Cardy formula
\cite{CARDY} (as most happen to do \cite{CAR3}).

A brief discussion on the methodology used in 
\cite{MMV,Prior} is also in order:  After establishing a suitable
coordinate system, we wrote out each of the metric components
as a Taylor expansion with respect to a Gaussian normal
coordinate (measuring distance from the horizon).
We then constrained the expansion coefficients  by
enforcing  regularity on geometric invariants as the horizon is
approached.  Once all possible constraints have been exhausted, it is
  straightforward to calculate the Einstein tensor to any order in the 
perturbative expansion. 

Although technically sound, our previously used method
does have --- arguably --- one possible shortcoming; namely, the static
case and the rotating (but stationary) case have to be treated separately.
By this, it is meant that, before proceeding with the calculation, one
must first  specify
a particular coordinate system; with such a choice being 
unavoidably different
for the  two stated cases. Meanwhile, it is a common school of thought
that any profound statements about gravity
should be expressible (at least in principle) in a
background or coordinate independent
framework \cite{ROV}.
With this philosophy in mind,
it has been our contention that there should be some
geometrical argument  which would confirm these
on-horizon symmetries, but without resorting to case
or coordinate specifics.
In fact, the beginnings of such an argument were laid out
in \cite{Prior}, although we were unable to
proceed past an intuitive level   at that time. 
In  the current paper we are, however,
able to alleviate  this situation and    establish
the validity of the argument --- up to a ``loophole'', which is elaborated
upon in the final section.

The rest of this paper proceeds as follows.
In the next section, we clarify our objective
while introducing important notation. The main calculations
in support of the argument are then presented in Section III.
Finally, there is a  summary
and some further discussion in the concluding section.

\section{Some Preliminaries}

To reiterate, it is our intention to show that,
for {\it any}  stationary (non-extremal) black hole spacetime,
the  following will occur {\it at the horizon}:
the Einstein tensor takes on a block-diagonal form (separating into
transverse and parallel blocks) and  the transverse
components of this tensor are directly proportional to the 
(induced) transverse metric. This is to be accomplished by a purely
geometrical argument that, unlike the  prior  method \cite{MMV,Prior},
does not have to be reformulated in going from  the static to the rotating
case (or {\it vice versa}).

Our argument will essentially depend on only three inputs: {\it (i)}
 the well-established properties of a stationary  and 
non-extremal Killing horizon,~\footnote{See \cite{Wald} for a
text-book discussion on Killing horizons. It should be noted, in particular, 
that
stationarity ensures that the horizon surface gravity is a constant \cite{Racz}
({\it i.e.}, the ``zeroth law'' of black hole mechanics),   while
non-extremality necessitates that this constant is non-vanishing
(essentially, the ``third law'').} 
{\it (ii)} time-reversal symmetry, and
{\it (iii)} the conservation of the Einstein tensor.
We will also ``assume'' a bifurcate Killing horizon; that is, one
that contains a bifurcation surface or spacelike 2-surface 
where the Killing vector vanishes. However, thanks to 
the work of  Racz and Wald \cite{Racz2},
the presence of such a bifurcation surface can be viewed as a consequence
of the first input rather than an additional assumption. To elaborate,
these authors demonstrated that, if the surface gravity is
constant and non-vanishing over any spacelike cross-section of a Killing
horizon,   there will exist a stationary extension of
the spacetime that does include a well-defined bifurcation surface.
Which is to say,  even  a {\it physically relevant} black hole 
(such as one formed
by stellar collapse) will asymptotically approach this type
of spacetime.

Before discussing the Einstein tensor, we
will first require a suitable basis for the on-horizon
coordinate system.~\footnote{This step
may appear to be contradictory with our earlier comment
about coordinate independence; however, see Section IV for why this is
not really so.}  
For this purpose, let us consider an arbitrarily chosen 
spacelike section of the Killing horizon. We will use $[m_1]^a$ and $[m_2]^a$
to denote  any orthonormal  pair of  spacelike vectors that 
are tangent to
this 2-surface.
 Then a convenient choice of basis  turns out to
be \cite{Prior} the orthonormal set  $\{\chi^a$, $N^a$, $[m_1]^a$, $[m_2]^a\}$;
where $\chi^a$ denotes the  Killing vector (which is null on the horizon),
 and  $N^a$
represents a vector that is both null on the horizon and 
orthogonal to $\chi^a$.
Note that the vector  $N^a$ is guaranteed to exist \cite{Wald}, although
it can only be fixed up
to a constant factor. Here, this constant will be specified (without loss 
of generality)  by  the normalization condition
 $\chi^a \;N_a = -1\;$.  Further note that --- except
for   $[m_1]^a\;[m_1]_a= 1$ and  $[m_2]^a\;[m_2]_a = 1$ ---
any other contraction of basis vectors must  vanish  on the horizon.


Since the prescribed set forms, by construction, an orthonormal basis for
the tangent space, we can express the on-horizon Einstein tensor 
(indeed, any symmetric on-horizon tensor) in the following pedantic manner:
\begin{eqnarray}
G_{ab} &=& \; G_{++} \;\chi_a \chi_b \;+\;  G_{--} \; N_a N_b 
\;+\; G_{+-} \;\left\{ \chi_a N_b + N_a \chi_b \right\}
\nonumber \\ 
&&
\;+\; G_{+1} \;\left\{ \chi_a [m_1]_b + [m_1]_a \chi_b \right\}
\;+\; G_{+2} \;\left\{ \chi_a [m_2]_b + [m_2]_a \chi_b \right\}
\nonumber \\
&&
\;+\; G_{-1} \;\left\{ N_a [m_1]_b + [m_1]_a N_b \right\}
\;+\; G_{-2} \;\left\{ N_a [m_2]_b + [m_2]_a N_b \right\}
\nonumber \\
&&
\;+\; G_{11} \; [m_1]_a [m_1]_b \;+\;  G_{22} \; [m_2]_a [m_2]_b
\nonumber \\
&&
\;+\; G_{12}  \; \left\{ [m_1]_a [m_2]_b + [m_2]_a [m_1]_b \right\}
\;,
\label{ET}
\end{eqnarray}
where the coefficients are, at this point, arbitrary
functions of the spacetime geometry.~\footnote{However, even at
this stage, we do know the following since
$\chi^a$ is a Killing vector:
any of these coefficients vanishes when operated on by $\chi^a\nabla_a\;$.}
By an inspection of equation (7.1.15) from Wald's  textbook \cite{Wald},
it is not difficult to verify that 
$G^b{}_a \; \chi_b \propto \chi_a\;$ must be true on the horizon \cite{Prior}.
(Meaning that the on-horizon Einstein tensor possesses a null 
eigenvector.) With knowledge of this fact (along with the 
orthogonality properties of the basis set), 
we can first contract equation (\ref{ET})  with $\chi_b$ and then 
set the coefficient to zero of any  term which is {\it not} 
proportional
to $\chi_a$. By way of this procedure, we have
$G_{--} = G_{-1} = G_{-2} = 0\;$, so that the on-horizon
form of the Einstein tensor reduces to 
\begin{equation}
G_{ab} = H_{ab} +J_{ab}\;,
\end{equation}
where
\begin{eqnarray}
H_{ab} &\equiv& \; G_{+-}\;  \left\{ \chi_a N_b + N_a \chi_b \right\}
\nonumber \\
&&
\;+\; G_{11} \; [m_1]_a [m_1]_b +  G_{22}\; [m_2]_a [m_2]_b
\nonumber \\
&&
\;+\; G_{12} \;   \left\{ [m_1]_a [m_2]_b + [m_2]_a [m_1]_b \right\} 
\label{HT}
\end{eqnarray}
and
\begin{eqnarray}
J_{ab} &\equiv& \;  G_{++} \; \chi_a \chi_b 
\nonumber \\
&&
\;+\; G_{+1} \; \left\{ \chi_a [m_1]_b + [m_1]_a \chi_b \right\}
\;+\; G_{+2} \; \left\{ \chi_a [m_2]_b + [m_2]_a \chi_b \right\} \;.
\label{JT}
\end{eqnarray}

We have separated $G_{ab}$ into two distinct tensors  so
as to readily distinguish between the part that survives
on the bifurcation surface --- where 
$\chi^{a}=0$~\footnote{Note, however, that
$\chi_a N_b$ has a well-defined limit on the bifurcation surface,
even though $\chi^a\rightarrow 0\;$. This is because 
the other null normal limits as  $N^{a}\rightarrow \infty\;$,
as a consequence of
the normalization condition $\chi^a \;N_a = -1\;$.}
--- and the 
remainder.  That is, by definition, 
$J_{ab}=0$ and   $G_{ab}=H_{ab}$ on the bifurcation surface. 
It is a further point of interest
that $H_{ab}$ can be also written in the following compact form:
\begin{eqnarray}
H_{ab} &=& \; G_{+-} \; [g_\perp]_{ab} \;+\; [H_\parallel]_{ab} \;,
\end{eqnarray}
where  
\begin{equation}
[g_\perp]_{ab}= \chi_a N_b + N_a \chi_b
\end{equation}
 is the transverse part of the  induced metric \cite{Wald}
and  we have defined the following ``in-horizon''  tensor: 
\begin{equation}
[H_{\parallel}]_{ab}\equiv \; G_{11} 
\; [m_1]_a [m_1]_b \;+\;  G_{22} \;[m_2]_a [m_2]_b
\;+\; G_{12}  \; \left\{ [m_1]_a [m_2]_b + [m_2]_a [m_1]_b \right\}\;. 
\end{equation}

It is clear from  the above 
formulation that  $H_{ab}$ is a block-diagonal tensor for which
the transverse components are proportional to the transverse metric.
Hence, it will be sufficient, for our purposes, to show that
$J_{ab}=0$ everywhere on the horizon. We will now set
out to  argue that this is indeed the case.

\section{The Main Argument}

Let us begin this section by focusing on  the tensor $H^{ab}\;$; that is,
 the
bifurcation-surface form of the Einstein tensor. Our initial objective
is to demonstrate that, when propagated {\it away} 
from the bifurcation surface but
{\it along} the horizon,  this tensor 
does not change
in relation to the metric (and, hence, in relation to
the spacetime geometry). Such a propagation is clearly a Killing translation,
and so it is significant that
 $H^{ab}$ is a Killing invariant
or ${\cal L}_{\chi}H^{ab}=0$ (which follows from
each individual constituent of $H^{ab}$ being a Killing 
invariant~\footnote{Note that ${\cal L}_{\chi}$ represents
a Lie differentiation  with respect to the Killing 
vector. See \cite{Wald} for some pertinent background.}). 
Consequently,
this tensor will be {\it formally} unaltered under any such
 translation along the horizon (as,  incidentally, so will  $J^{ab}$). 
It  is then sufficient for our purposes to establish the on-horizon
validity of 
$\chi^c\nabla_c H^{ab}=0\;$. The point being that,
if this is indeed correct,  $H^{ab}$ will be {\it parallel}
transported as it moves away from the bifurcation surface; meaning that
its relationship with the metric (which is certainly parallel transported)
will remain intact.

First of all, let us consider the transverse part of $H^{ab}$ or
$G_{+-}[g_\perp]^{ab}\;$. 
This is the product of a scalar coefficient 
and a block-diagonal part of the ``on-horizon metric''
({\it i.e.}, the projection of the spacetime metric onto the horizon).
As long as we are staying on the horizon ---  which is  implied by
the Killing translation ---  this induced metric (and, hence, any
block-diagonal constituent thereof) will also be invariant
under the action of $\chi^a\nabla_a\;$. Therefore, on the horizon,
\begin{equation}
\chi^c\nabla_c \{ G_{+-}\;[g_\perp]^{ab}\} =0\;.
\label{above}
\end{equation}

We are  now left with the task  of verifying the on-horizon validity of  
 $\chi^c\nabla_c [H_{\parallel}]^{ab}=0\;$.
To demonstrate this result, let us consider, in turn,
the following  two possibilities for the state of
a stationary black hole spacetime: either  the black hole is static
{\it or} it is axially symmetric (and possibly rotating)
\cite{Wald}. [The
essential point here is that, if a rotating black hole is embedded in
a spacetime that is {\it not} axially symmetric,  
tidal forces will act to both slow down the rotation and smooth 
out the asymmetry.  Hence, the spacetime will continue to
evolve until such time as either staticity or axial symmetry
has been achieved.] 

Firstly, if the black hole is static,  there must be
 a timelike Killing vector (say, [$\partial_t]^a$) for the spacetime. 
Moreover, the {\it on-horizon} action of the
 operator $\chi^a\nabla_a$ is equivalent to that of
[$\partial_t]^a\nabla_a$ and, therefore,
$\chi^c\nabla_c [H_{\parallel}]^{ab}=\nabla_{t} [H_{\parallel}]^{ab}\;$.
Since there can be no timelike components for the in-horizon
2-metric (say, $[g_{\parallel}]_{ab}$) in the static case, one
can readily verify that this covariant  derivative
reduces to
\begin{equation}
\nabla_{t} [H_{\parallel}]^{ab}\;\ =\;\ 
\partial_{t}[H_{\parallel}]^{ab} \; + \;
{1\over 2}  [H_{\parallel}]^{ac} [g_{\parallel}]^{db}
\;\partial_{t}[g_{\parallel}]_{cd}
 \; + \;
{1\over 2}  [H_{\parallel}]^{cb} [g_{\parallel}]^{ad}
\;\partial_{t}[g_{\parallel}]_{cd}\;,
\end{equation}
which is, of course, trivially vanishing.  

Secondly, if the black hole is axially symmetric, then the
axis of rotation picks out a particular spacelike direction
(say, $\phi$), which is significant for the following reason:
Given our  ``input'' of time-reversal
symmetry,
it follows that every physically relevant quantity should be invariant
under the simultaneous change of $t\rightarrow -t$ and
$\phi\rightarrow -\phi\;$. In view of this fact and  
$[\partial_\phi]^a$ being a Killing vector \cite{Wald},  
if  we choose (without loss
of generality)  to orientate $[m_1]$
in the $\phi$ direction, there can be no off-diagonal elements
in $[H_{\parallel}]^{ab}\;$.  
By similar reasoning,  the  in-horizon 2-metric ($[g_{\parallel}]^{ab}$)
will also have no off-diagonal elements.
And so we can, after  diagonalizing both tensors, arrive at the form  
\begin{equation}
[H_{\parallel}]_{ab} = {\tilde G}_{11} 
\; [g_{m_1}]_{ab} \;+\; {\tilde G}_{22} \; [g_{m_2}]_{ab}\;,
\end{equation}
where  the  tilde on
 a coefficient  signifies a suitable redefinition
and the metric ``blocks'' have been labeled accordingly.
Since each term in  $[H_{\parallel}]_{ab}$ is the product
of a scalar and a block-diagonal constituent of the metric, 
it can be deduced  [recalling the
discussion leading up to equation (\ref{above})] 
that~\footnote{For the sceptical reader, there is
another way of seeing the very  same thing. Let us begin with 
${\cal L}_\chi [m_{1}]^a [m_{2}]^b=[m_{1}]^a\chi^c\nabla_c[m_{1}]^b +
[m_{1}]^b\chi^c\nabla_c[m_{1}]^a -[m_{1}]^a [m_{1}]^c\nabla_c\chi^b
-[m_{1}]^b [m_{1}]^c\nabla_c\chi^a = 0\;$. Next, we will
 contract this Lie derivative   
with the (diagonalized) 
on-horizon metric  or $g_{ab}=[g_{\perp}]_{ab}+g_{11}[m_1]_a[m_1]_b
+g_{22}[m_2]_a[m_2]_b\;$. This process yields 
$2g_{11}[m_1]_a\chi^{b}\nabla_b[m_1]^a
- 2g_{11}[m_1]_a[m_1]^b\nabla_b\chi^a=0\;$.  The antisymmetry property
$\nabla_a\chi_b=-\nabla_b\chi_a$ \cite{Wald} 
means that the second term  vanishes,
which then implies that $\chi^{c}\nabla_c[m_1]_a[m_1]^a=
g_{ab}\chi^{c}\nabla_c[m_1]^a[m_1]^b=0\;$.
This outcome (and the $[m_2]$ analogue) and the diagonality
of $[H_{\parallel}]^{ab}$ is sufficient  to reproduce equation 
(\ref{check}). Moreover, the same  argument substantiates
equation (\ref{above}), since the transverse metric can always be 
recast in a diagonal form.} 
\begin{equation} 
 \chi^c\nabla_c [H_{\parallel}]^{ab}=0\;
\label{check}
\end{equation}
is, once again, true on the 
horizon. (Note that, even though a distinction was made, 
the end  result      
is a coordinate-independent statement.)

Let us now bring  the conservation  of the (on-horizon)  
Einstein tensor into
the game. More specifically, we will use 
\begin{equation}
\nabla_a G^{ab} =\nabla_a H^{ab} + \nabla_a J^{ab} \rightarrow 0 \; ;
\label{problem}
\end{equation}
where the arrow signifies that the resulting derivatives have
been suitably ``pulled back'' to the horizon.
As implied by this last remark, we must now deal with an important caveat:
In general, the action of these covariant derivatives
will move the tensors off of the horizon (albeit, infinitesimally),
and it is no longer obvious how to handle computations involving
the Killing vector.  
One  way around this dilemma
is to ``replace'' the  Killing vector
$\chi^a$ with a vector $\rho^a$ which is defined by
\cite{CARX} 
\begin{equation}
\nabla_{a} |\chi|^2 \;=\; -2\kappa\rho_a \;.
\label{definition}
\end{equation}
(Here,  $\kappa$ is the surface gravity ---
which can itself  be defined by the relation  
$\chi^a\nabla_a \chi^b=\kappa \chi^b$ on the horizon \cite{Wald}.)
The relevant point being that $\rho^a$ has been  precisely defined
so that it {\it always} satisfies $\rho^a\; \chi_a=0$ and
 becomes null on the horizon; that is, $\rho^a$ limits
to $\chi^a$ at the horizon \cite{CARX}. 
Hence, in obtaining any on-horizon result, 
we   can first do the 
calculation in terms of $\rho^a$ and then take the limit
to the horizon afterwards.~\footnote{Technically speaking,
a similar replacement should be made for $N^a\;$. 
This is rather tricky, given our lack of knowledge about 
this vector under general circumstances. 
Fortunately, the matter never explicitly comes up.
However, see Section IV for some discussion on  a related subtlety;
that is, the ``loophole''.}  

Let us begin here with the observation that, on the bifurcation surface
in particular,
$\nabla_a J^{ab} \rightarrow 0 \;$.
To see this, it is useful to call upon a few on-horizon limits:
$\nabla_a\rho^a \rightarrow 2\kappa$ 
[{\it cf}, \cite{CARX}; equations (A.7) and (A.10)], 
$\rho^a \nabla_a \rho^b \rightarrow \kappa \chi^b$
 [{\it cf}, \cite{CARX}; equations (A.8)
and (A.10)],
and
$[m_1]^a\nabla_a \rho^b \rightarrow [m_1]^a\nabla_a \chi^b 
$ (and similarly for $[m_2]^a$).~\footnote{These limits 
can be obtained by long but straightforward 
calculations \cite{CARX}
that incorporate the defining relation  (\ref{definition})  for $\rho^a\;$,
and the definition 
of the surface gravity  as the horizon limit of the
Killing orbit acceleration or $\sqrt{\nabla^a|\chi|\nabla_a|\chi|}$
\cite{Wald}.} 
By application of these  limiting relations, it is not difficult to determine
from equation (\ref{JT})
what the ``residue'' of  $\nabla_a J^{ab}$ turns out to be 
(after pulling back to the
bifurcation surface,
where $J^{ab}$ is itself identically vanishing). Given that
$\nabla_a\rho^a \rightarrow 2\kappa$  whereas (on the horizon) 
$\nabla_a\chi^a = 0 \;$, the residue in question is just
$\;2\kappa \{G_{++}\chi^b + G_{+1}[m_1]^b +G_{+2}[m_2]^b\}\;$. 
Now,  from this last expression, we can see that the residue
is, in fact,  directly proportional to  
the on-horizon contraction 
of $N_{a}$ with $J^{ab}\;$. Since this contraction must certainly vanish on the
bifurcation surface,    
$\nabla_a J^{ab} \rightarrow 0$ then follows.

 The above outcome and equation (\ref{problem}) allows
us to deduce that
 $\nabla_a H^{ab}\rightarrow 0$ on the bifurcation surface.
But actually, since
$H^{ab}$ is now known to maintain its formal relationship with 
the geometry
 when 
propagated in the Killing direction  ({\it cf}, the beginning
of the section), 
 this tensor should be similarly conserved 
at {\it any} point along the horizon.
It  immediately follows that, anywhere on the horizon,
\begin{equation}
\nabla_a J^{ab} \rightarrow 0
\label{CJ} 
\end{equation}
should be  a true statement.

Now substituting equation (\ref{JT}) into equation (\ref{CJ}),
we find that
\begin{eqnarray}
0 &=& \; 3\kappa G_{++}\;\chi^b  
\nonumber \\
&&
\;+\; G_{+1}\;\{2\kappa [m_1]^b + \chi^a\nabla_a [m_1]^b
+[m_1]^a\nabla_a\chi^b + \chi^b\nabla_a[m_1]^a\}
\nonumber  \\
&&
\;+\; G_{+2}\;\{2\kappa [m_2]^b + \chi^a\nabla_a [m_2]^b
+[m_2]^a\nabla_a\chi^b + \chi^b\nabla_a[m_2]^a\}
\nonumber \\
&&
\;+\;\{[m_1]^a\nabla_a G_{+1}
+[m_2]^a\nabla_a G_{+2}\}\;\chi^b \;,
\label{YYY}
\end{eqnarray}
where --- once again --- the calculation is  done initially
 in terms of $\rho^a$
and then the on-horizon limit is taken. 

Next, let us contract  the right-hand side with $[m_1]_b\;$, 
which then yields
\begin{equation}
0= \; 2\kappa G_{+1} \;+\; G_{+2}\;\{[m_1]_b\chi^a\nabla_a[m_2]^b
 \; +\; [m_1]^b[m_2]^a\nabla_a \chi_b \}\;.
\label{XXX} 
\end{equation}
Here, we have used the various orthogonality properties,
as well as  ${\cal L}_{\chi}[m_1]^b= 
\chi^a\nabla_a [m_1]^b -[m_1]^a\nabla_a\chi^b=0\;$.
Doing the same with $[m_2]_b\;$, we similarly obtain
\begin{equation}
0=\; 2\kappa G_{+2} \;+\; G_{+1}\;\{ [m_2]_b\chi^a\nabla_a[m_1]^b
  \;+\; [m_2]^b[m_1]^a\nabla_a \chi_b \}\;.
\label{YYYY} 
\end{equation}
But actually, if we apply $[m_2]_b\; [m_1]^b=0$
to the first term in the curly brackets and $\nabla_a\chi_b=-\nabla_b\chi_a$
to the second, this last equation can be recast as
\begin{equation}
0= \;2\kappa G_{+2} \;-\; G_{+1}\;\{ [m_1]_b\chi^a\nabla_a[m_2]^b
  \;+\; [m_1]^b[m_2]^a\nabla_a \chi_b \}\;.
\label{YYYYY} 
\end{equation}

The above computations (\ref{XXX},\ref{YYYYY}) 
may be reinterpreted as a system of two equations 
 with two unknowns
($G_{+1}$, $G_{+2}$); that is,
\begin{eqnarray}
0 &=& 2\kappa G_{+1} \;\ +\;\ {\cal Z} G_{+2}
\nonumber \\
0 &=& 2\kappa G_{+2} \;\ -\;\ {\cal Z} G_{+1} \;,
\end{eqnarray}
where 
${\cal Z}$ is just the contents of the curly brackets.
The requirement for a non-trivial solution turns out to be
$4\kappa^2 =-{\cal Z}^2\;$.
But,
since $\kappa >0$ by hypothesis (and ${\cal Z}$ is clearly real), 
this must mean that
$G_{+1}= G_{+2}=0\;$.  Consequently [{\it cf},
equation (\ref{YYY})], $G_{++}=0\;$. The end result
of all this is that  $J_{ab}=0$ [{\it cf}, equation (\ref{JT})],
and so $G_{ab}=H_{ab}\;$.  Which is to say,
the Einstein tensor takes on the highly symmetric
(block-diagonal) form
\begin{equation}
G_{ab} \;=\; G_{+-} \; [g_\perp]_{ab} \;+\; [H_\parallel]_{ab} \;,
\end{equation}  
{\it anywhere} on the horizon.

\section{A Discussion}

In summary, we have utilized a particularly elegant method
to demonstrate   the highly symmetric nature   of the
Einstein tensor near a Killing horizon.  More specifically, we
have shown that,
for any stationary (non-extremal) Killing horizon,
the on-horizon Einstein tensor block-diagonalizes into transverse
and parallel blocks. Additionally,  the transverse block
is constrained to be
directly proportional to the transverse metric.

It is noteworthy that (along with M. Visser and D. Martin) we 
were able to deduce the very  same
near-horizon symmetries  by a substantially different approach 
\cite{MMV,Prior}.
These earlier treatments were based on first expanding
out the metric components (in terms of normal distance from the horizon) 
and then enforcing regularity on the  near-horizon geometry.
In spite of the apparent duplicity,
the current work has a distinct  advantage over the former: 
it has allowed us to handle both relevant cases (static and  rotating)
with  a single iteration. This simplifying feature can
be attributed to the use of an essentially
coordinate-independent approach. To see the validity of this statement, 
consider
that the  Killing vector, $\chi^a\;$, can be interpreted as a purely geometric 
entity.  Given this interpretation, the other three vectors in our basis 
($N^a$, $[m_1]^a$, $[m_2]^a$) 
can be suitably defined and then  viewed as relational constructs 
with respect to $\chi^a\;$. 
Hence, the analysis really required  no formal coordinate system in the usual
sense. 

The attentive reader will notice that the word ``proof'' was never 
used in the main text. This is because our analysis cannot be regarded
as rigorous in the following sense: By our use of
the conservation equation (\ref{problem}), it
was necessary (at least implicitly) to extrapolate the tensors 
$H_{ab}$ and $J_{ab}$ {\it away} from the horizon.
It was, however, never actually confirmed that these tensors
maintain the distinction of being {\it separate} entities
under such an extrapolation. [Note,
though, that
the total Einstein tensor or $G_{ab}$ will {\it always} be a
well-defined object --- both on and outside of the horizon --- irrespective 
of the status of $H_{ab}$ and $J_{ab}$ individually.]
Hence, it can  not be said with {\it absolute}  certainty
 that $H^{ab}$ and $J^{ab}$
will be separately conserved, in spite of the compelling arguments of 
Section III.
 
Unfortunately, our ability to  resolve the above issue 
is hindered
by a lack of knowledge --- under {\it general} circumstances ---
about the off-horizon behavior of the vector $N^{a}\;$.  On a more
favorable note,  the results
of the current derivation do happen to agree with those of our prior and
{\it completely independent} work \cite{MMV,Prior}.
Nonetheless, this ``loophole'' is rather bothersome, and
we hope  to (somehow) rigorously address this  issue
at a later time.

Let us re-emphasize that this highly symmetric form
of Einstein tensor implies (via the field equations)
that the stress tensor near a black hole horizon
will be similarly constrained. This can be expected to
have severe repercussions on the matter
and energy --- including any quantum fluctuations ---
that can exist  near a horizon. As discussed 
and elaborated on elsewhere \cite{MMV,Prior},
we suspect that the symmetries at hand may
be relevant to some recent (statistically based) calculations of
the black hole entropy; most notably, those of  Carlip 
\cite{CAR1,CARX}.
It remains an ongoing challenge, however, to place this connection
on firmer, more rigorous ground. (But, for some first steps
in this direction, see \cite{CVI}.)

Another interesting challenge will be to generalize our findings to
extremal Killing horizons, as well as to  dynamical 
spacetimes. [Yet, if the spacetime is evolving slow
enough (or is ``quasi-stationary''), 
then our results should still be applicable
in an approximate sense.]
Alas, the extremal case is plagued by conceptual issues,
whereas truly dynamical scenarios present 
difficulties of a more technical nature.
Suffice it to say, such matters (like the previous ones) 
are currently under investigation.

\newpage

\section*{Acknowledgements}

The author thanks Matt Visser for his instructive guidance and insightful 
criticisms,
even though we were not  always {\it simpatico}.
The author also thanks Damien Martin for earlier inputs and Steve Carlip
for asking a motivating  question at ``KerrFest''.
Research supported by the Marsden Fund administered by the Royal
Society of New Zealand, and by the University Research Fund of
Victoria University.



\end{document}